\documentclass[12pt]{article}
\usepackage{amssymb}
\usepackage[usenames,dvipsnames]{color}
\usepackage{hyperref}
\usepackage[left= 1in,right=1in,top=1in,bottom=1in]{geometry} 

\hypersetup{
    colorlinks=true,         
    linkcolor=blue,          
    citecolor=red,        
    urlcolor=Violet             
}

\newcommand{\f}{\begin{equation}}
\newcommand{\ff}{\end{equation}}

\title{A Shape Dynamical Approach to Holographic Renormalization}
\author{Henrique~Gomes,$^1$
Sean~Gryb,$^{2,3}$
Tim~Koslowski,$^4$\\
Flavio~Mercati$^5$ and Lee Smolin$^5$
\footnote{electronic addresses: {s.gryb@hef.ru.nl}, {gomes.ha@gmail.com,} {t.a.koslowski@gmail.com},  {fmercati@perimeterinstitute.ca}, {lsmolin@perimeterinstitute.ca}.}
 \vspace{12pt} \\
\it \small $^1$University of California at Davis, One Shields Avenue Davis, CA, 95616, USA,\\
\it \small $^2$Institute for Theoretical Physics, Utrecht University, \\
\it \small Leuvenlaan 4, 3584 CE Utrecht, The Netherlands,\\
\it \small $^{3}$Radboud University Nijmegen, Institute for Mathematics, Astrophysics and Particle Physics, \\
\it \small  $^4$ University of New Brunswick, Fredericton, NB, E3B 5A3 Canada,\\
\it \small $^5$Perimeter Institute for Theoretical Physics, 31 Caroline Street North,\\
\it \small Waterloo, ON, N2L 2Y5 Canada.}

\date{\today}

\begin{document}

\maketitle

\begin{abstract}
We provide a bottom-up argument to derive some known results from holographic renormalization using the classical bulk-bulk equivalence of General Relativity and Shape Dynamics, a theory with spatial conformal (Weyl) invariance. The purpose of this paper is twofold: 1) to advertise the simple classical mechanism: trading of gauge symmetries, that underlies the bulk-bulk equivalence of General Relativity and Shape Dynamics to readers interested in dualities of the type of AdS/CFT; and 2) to highlight that this mechanism can be used to explain certain results of holographic renormalization, providing an alternative to the AdS/CFT conjecture for these cases. To make contact with usual the semiclassical AdS/CFT correspondence, we provide, in addition, a heuristic argument that makes it plausible why the classical equivalence between General Relativity and Shape Dynamics turns into a duality between radial evolution in gravity and the renormalization group flow of a conformal field theory. We believe that Shape Dynamics provides a new perspective on gravity by giving conformal structure a primary role within the theory. It is hoped that this work provides the first steps towards understanding what this new perspective may be able to teach us about holographic dualities.
\end{abstract}

\newpage
\tableofcontents
\newpage

\section{Introduction}\label{sec:Intro}

A key insight of holography is that the behaviour of a conformal field theory (CFT) under renormalization is governed by a diffeomorphism-invariant gravitational theory in one higher spacetime dimension.  In this correspondence, the extra dimension plays the role of the renormalization scale.  This gives the renormalization group a geometric setting and suggests deep connections between the dynamics of spacetime and the renormalization of quantum fields. 

The holographic formulation of the renormalization group was elucidated in a beautiful series of papers \cite{Skenderis:holo_weyl,Verlinde,Skenderis:holo_RG,Skenderis:lecture_notes,Skenderis:Holo_ren_review,Skenderis:how_to_go,Skenderis:holoRG_main,McFadden:holo_cosmo,Freidel}  and underlies much of the very fertile applications of holography and the AdS/CFT correspondence \cite{Maldacena,Witten,Gubser:adscft} to condensed matter and fluid systems.  In this paper, we offer new insights into the connection between gravitation and renormalization by giving a new explanation for why that connection exists.  
We give a novel derivation of a correspondence between a CFT defined on a $d$-dimensional manifold $\Sigma$ with metric $G_{ab}$, 
and a solution to a gravitational theory in $d+1$ dimensions. In this
correspondence, $\Sigma$ is to be interpreted as the asymptotic boundary of a Euclidean
asymptotically $AdS$ spacetime, $\cal M$, the latter being a solution to a spacetime
diffeomorphism invariant theory of a metric $g_{\mu \nu}$ such that $\Sigma = \partial {\cal M}$ and $G_{ab}$ is
the pullback of $g_{\mu \nu}$  onto $\Sigma$.\footnote{We will highlight the differences and similarities between ours and the usual holographic renormalization program when they arise.} 
We give a plausibility argument for the conjecture that this correspondence can be seen as the reason why gravitational evolution encodes the renormalization group flow of a  $CFT$, at least to leading  order in a semiclassical regime, near the conformal boundary of a Euclidean locally asymptotically AdS spacetime. 

The new derivation we offer makes use of a recent reformulation of General Relativity in which spatial conformal invariance plays a central role.  This formulation, called Shape Dynamics, is the result of trading the many fingered time aspect of spacetime diffeomorphism invariance --- which allows one to arbitrarily redefine what is meant by surfaces of constant time --- with local Weyl transformations on a fixed class of spatial surfaces.  In a word,
Shape Dynamics trades relativity of time for relativity of scale. Each is a gauge invariance that can be defined on the phase space of general
relativity depending on one free function. The purpose of this paper is not only  to re-derive some known results through a novel route, but to highlight a mechanism, which is an alternative to the AdS/CFT conjecture, that can be used to explain some results that are usually attributed to the AdS/CFT conjecture. 
This mechanism is the trading of classical gauge symmetries, which we will explain in section \ref{sec:Mechanism}. The purpose of this paper is thus to show that this mechanism is very general and that this general mechanism can be used to derive known results without using the AdS/CFT conjecture. An aspiration of the current work is that it lay down the foundations for exploring further interesting connections between Shape Dynamics and holography.

\subsection{Heuristics}

Before we posit a particular correspondence between a CFT and a particular gravitational theory in one higher dimension, we can ask a more general question:  what properties do generic gravitational theories and generic CFT's share which allows us to posit correspondences between them?   One well-known answer builds on global symmetries, in particular, the correspondence between the AdS group in $d+1$ dimensions and the global conformal group in $d$ dimensions.  Here we would like to suggest that a deeper correspondence holds between gauge symmetries, in particular between the group of refoliations of the spacetime manifold in $d+1$ dimensions and the group of Weyl transformations acting to locally rescale the metric on its boundary.  This connection between refoliations and Weyl transformations is captured by the mechanism of gauge symmetry trading which is at the heart  of Shape Dynamics.  The main idea we want to develop in this paper is that it is the existence of gauge symmetry trading between 
refoliations of a $d+1$ dimensional spacetime and Weyl transformations on  fixed foliations of that spacetime that provide a deep and very general reason why there exist correspondences between bulk gravitational theories and boundary CFTs\footnote{Note that it is already known that if a classical field theory is Weyl invariant for a general metric, $h_{ab}$ the theory will be invariant under global conformal transformations when that metric is taken to
be flat~\cite{Faci:weyltoconformal}.}.

The main work of this paper consists of showing how trading of spacetime refoliation invariance for spatial Weyl invariance can be used to explain a subset (which will be explicitly specified later) of the results of holographic renormalization. This trading of spacetime refoliation invariance for spatial Weyl invariance has some immediate consequences for usual CFTs, since invariance of a field theory under Weyl transformations for a general metric implies invariance under the global conformal group when that metric is (conformally) flat \cite{Faci:weyltoconformal}.

We will see that it is very useful to consider separately constant scalings that affect the total spatial volume, and to distinguish these from Weyl transformations that leave the spatial volume fixed.  As we will see, these volume preserving conformal transformations (VPCT) play a special role in Shape Dynamics.

\subsection{Shape Dynamics}\label{sec:SD}

The main argument of this paper is based on number of observations made during the development of the Shape Dynamics description of General Relativity, which was in part used in \cite{gryb:gravity_cft}. These observations are:
\begin{enumerate}
 \item There is a classical mechanism ({\it gauge symmetry trading}), based on two partial gauge fixings of a {\it linking gauge theory}, which generates exact dualities between classical gauge theories \cite{Gomes:linking_paper}.
 \item There is a construction principle for gauge symmetry trading based on the generalized {\it St\"uckelberg mechanism}, also called {\it Kretschmannization}, by relativists \cite{gryb:shape_dyn}.
 \item The application of this mechanism to classical General Relativity results in a theory in which the gauge symmetry due to refoliations of spacetime is traded for a gauge symmetry of spatial Weyl--transformations that preserve the total spatial volume of a compact Cauchy slice (VPCT\footnote{VPCT stands for (total) Volume Preserving Conformal Transformations.} symmetry). This results in {\it Shape Dynamics}, which describes gravity as a dynamical theory of spatial conformal geometry \cite{Gomes:linking_paper,gryb:shape_dyn}.
 \item The physical predictions of the two descriptions of classical gravity, General Relativity and Shape Dynamics, are locally\footnote{Despite the exact local equivalence, there is a possibility for global differences between General Relativity and Shape Dynamics, if CMC gauge is not globally available.} equivalent, i.e. the solutions to the equations of motion differ only by gauge transformations. 
 The solutions coincide manifestly when General Relativity is evolved in constant mean (extrinsic) curvature (CMC) gauge and if Shape Dynamics is evolved in a conformal gauge determined by the Lichnerowicz--York equation \cite{York:york_method_prl}.
 \item General Relativity in CMC gauge is a {\it dictionary}, which relates observables of General Relativity  uniquely to observables of Shape Dynamics and vice versa. The local availability of CMC gauge allows one to translate {\it all} local physical statements from either theory into the other \cite{Koslowski:obs_equiv}.
 \item\label{obs:constantR} The homogeneous lapse equations of motion of General Relativity are the CMC equations of motion in any CMC slice in which the spatial Ricci scalar is homogeneous. It follows from the manifest coincidence with Shape Dynamics that the proper time equations of motion manifestly coincide with the equations of motion of a VPCT gauge theory \cite{Koslowski:SD_EFT}.
\end{enumerate}
A more detailed discussions of these observations follows in section \ref{sec:Mechanism}. 

\subsection{Main Results}\label{sec:Mainresults}


In this paper we consider some implications of gauge symmetry trading for the class of spacetimes which provide the setting for the AdS/CFT correspondence, which are called asymptotically locally AdS spaces.  Our main result is that we will explain, without referring to the AdS/CFT conjecture or any particular dual pairing of theories, why the $\pi^{ab}$ of a spacetime diffeomorphism invariant theory of a metric in the bulk has, when evaluated on the boundary of an $AlAdS$ spacetime,  several of the properties expected of a renormalized $T^{ab}$ of a CFT. The novel feature of the derivation we will give is that the \emph{integral} of the trace of $T^{ab}$ must be a Weyl-invariant functional of $G_{ab}$, which is a necessary (but not sufficient) condition for it to represent a conformal anomaly. In our derivation, this is a direct result of the local symmetry principle of Shape Dynamics, while in standard holographic renormalization calculations this result is purely coincidental unless one accepts the broader validity of the AdS/CFT conjecture.

Our result will be obtained in a series  of steps:
\begin{itemize}

 \item We explain in section \ref{sec:AlAdSandVPCT} that Euclidean asymptotically locally AdS (AlAdS) boundary conditions imply 
that the $r=const.$ slices\footnote{Where the radial coordinate is defined by equation (\ref{reef}), below.}
are asymptotically CMC and possess asymptotically vanishing spatial curvature when the conformal boundary is approached. It follows (from an argument analogous to observation \ref{obs:constantR}) that the radial evolution equations of General Relativity coincide manifestly with the equations of motion of a VPCT--gauge theory 
at the conformal boundary.  This provides the classical, geometric setting for the occurrence of a dual CFT defined on the boundary.

 \item  In section \ref{sec:HoloRen}  we derive the large volume asymptotic behavior of the Shape Dynamics Hamilton--Jacobi functional and recover a number of results that are obtained in the holographic renormalization framework, in particular an expression for the 
 integral of the holographic trace anomaly.  
 This is the finite piece of the integral of the trace of the metric's canonical momentum, ${\pi}$, evaluated at the boundary and it is the quantity that maps to the integral of the trace anomaly of the CFT in the AdS/CFT correspondence.  

 \item{} There is however an important subtlety to point out. The manifest coincidence of Shape Dynamics and General Relativity requires us to use the boundary limit of the induced metric, which will be denoted by $G_{ab}$ below, rather than the rescaled induced metric, which we will denote by $\gamma^{(0)}_{ab}$ below, that is used in standard presentations of holographic renormalization. Now, the AlAdS boundary conditions on $h_{ab}$ --- the pullback of the spacetime metric onto foliations ---  imply fall--off conditions for curvature invariants derived from $G_{ab}$ that allow one to only distinguish the integrals of these curvature invariants. This has an important consequence: In even dimension, $d$,  CFT's have a trace anomaly of a local form \cite{Duff:anomaly_review,Deser}
 \f
 T = \left< 0 \right| T \left| 0 \right> = \sqrt{h} \, Q \, ,
 \label{local}
 \ff
 where $Q$ is a local scalar polynomial in $\frac{d}{2}$ curvature components. Since the results presented here are stated in terms of $G_{ab}$, one can not recover the anomaly $Q(x)$, but only the integral thereof. The local counter term action and the local conformal anomaly found in standard holographic renormalization are, on the other hand, expressed in terms of $\gamma^{(0)}_{ab}$. The restriction to $G_{ab}$ is thus the mechanism by which VPCT invariance is {\it consistent} with the equivalence of General Relativity and Shape Dynamics at the boundary. In particular the AlAdS boundary conditions imply that the integrated anomaly, i.e. $\int \pi_{SD}$, tends to a constant at the boundary and its dependence on the boundary metric provides a nontrivial consistency check with the holographic renormalization group results. Our direct calculations in section \ref{sec:HoloRen} verify $\left.\int \pi_{SD}\right.|_{bdy.}=\left.\int T\right._{bdy.}$. An interesting extension of this work would be to redo the derivation performed here in terms of the rescaled metric $\gamma^{(0)}_{ab}$ in order to be able to compare more directly with the calculation of the local anomaly in standard approaches to holographic renormalization. We note, however, that our results are completely consistent, where applicable, with known calculations.
\end{itemize}

This main result can be expressed in the familiar AdS/CFT language by assuming the existence of a partition function for Shape Dynamics in the bulk, which admits a recognizable\footnote{``Recognizable'' means that the Ward--identities take the form of classical constraints when  momenta are substituted for  functional derivatives.} semiclassical limit near the boundary. Using these assumptions we argue in section \ref{sec:Speculation} that:

\begin{itemize}

 \item The assumption of a semiclassical limit of Shape Dynamics near the boundary induces equations that can be recognized as the UV--limit of Wilsonian renormalization group equations for a CFT. It is thus not necessary to assume the AdS/CFT conjecture to explain the appearance of a CFT at the conformal boundary, rather the manifest classical equivalence of the radial evolution at the boundary with the Shape Dynamics evolution explains the appearance of a CFT near the boundary.\footnote{ Note that the correspondence between bulk gravity and boundary CFT is purely coincidental from the point of view of standard holographic renormalization arguments unless the AdS/CFT conjecture is envoked.}
 
\end{itemize}

The combination of the classical gauge symmetry trading mechanism with this assumption leads to a construction principle for dualities of AdS/CFT type.  
This recipe is (1) constructive and (2) very general. We thus expect that it can be applied in a wide variety of circumstances. The recipe is summarized in \ref{sec:recipe}. 

We also want to emphasize that our reasoning turns the usual AdS/CFT argument upside down: we use the mathematical bulk--bulk equivalence of the Shape Dynamics and the ADM description of classical gravity and explore what this exact classical equivalence implies for a partition function in the semiclassical regime. I.e., we use a bottom--up approach to explain some known AdS/CFT results, rather than assuming the validity of the AdS/CFT conjecture and deriving results 
top--down.

The appendix~\ref{app:matterfields} of this paper provides the first hints of how our main results could be extended to more complicated CFTs. There, we consider the effect of a scalar field on the large volume behavior of the Shape Dynamics Hamilton--Jacobi functional, where we find agreement with results from holographic renormalization.

Unless otherwise specified, we restrict ourselves to the case where both bulk and boundary metrics are Euclidean and the cosmological constant,
$\Lambda =  -\frac{d(d-1)}{\ell^2} $ is negative. 

\subsubsection*{Notation}

Due to the usage of metrics of different dimensions, particular metric expansions, pull-backs etc, might prove slightly confusing. We thus provide the reader with a glossary of notation: 
\begin{eqnarray*}
g_{\mu\nu}~&:& ~~\mbox{The space-time metric on the locally asymptotically AdS manifold.}\\
h_{ab}~&:& ~~\mbox{The pull-back of the space-time metric onto the hypersurfaces of the foliation.}\\
\gamma_{ab}~&:& ~~\mbox{The rescaled hypersurface metrics: $r^2\gamma_{ab}=h_{ab}$.}\\
G_{ab}~&:& ~~\mbox{The asymptotic boundary value of $h_{ab}$ (as the radius goes to infinity).  }\\
\gamma^{(n)}_{ab}~&:&~~\mbox{The inverse radial even expansion of $\gamma_{ab}$, $\gamma_{ab}=\sum \gamma^{(n)}_{ab}r^{-2n}$}\\
^{(d+1)}R_{abcd}~&:& ~~\mbox{The full Riemmann tensor restricted to spatial indices.   }\\
~^{(d)}R_{abcd}~&:& ~~\mbox{ The foliation induced hypersurface intrinsic Riemann tensor.}
\end{eqnarray*}

\section{Mechanism}\label{sec:Mechanism}

This section serves as an introduction to those results about Shape Dynamics that are relevant for the construction of exact classical dualities in the present paper. The presentation is intended to highlight: (1) the simplicity and (2) the constructive nature of our mechanism. 

We will be interested in applying the gauge symmetry trading formalism to the ADM formulation of gravity in d+1 dimensions. This formalism is, however, very general and can be applied, not only to the case we are interested in here, but also to other familiar models. The power of the formalism is perhaps illustrated in this paper by the fact that, in the context of our derivation, it constitutes the key tool which explains the relation between gravity and renormalization of a CFT. The general formalism of gauge symmetry trading was introduced in \cite{gryb:shape_dyn} and formalized in \cite{Gomes:linking_paper}. The quantum aspects of these dualities were developed in \cite{Koslowski:obs_equiv} and some simple examples are discussed in \cite{Gomes:FAQs}. For convenience, we have included a technical appendix~\ref{app:sym_trading} which summarizes the general results.

Returning to the case of gravity, we begin with a heuristic motivation of gauge symmetry trading in General Relativity, then provide the technical details of the procedure.

\subsection{The Basic Idea of Gauge Symmetry Trading in General Relativity}

The Hamiltonian framework for General Relativity is based on an initial phase space, $\Gamma$, given by $d$ dimensional metrics, $h_{ab}$, on a manifold, $\Sigma$, and its  conjugate momenta, $\pi^{ab}$.  Physical solutions live on a constraint surface ${\cal C} \in \Gamma$, given by two kinds of first class constraints, ${\rm Diff}[v]=\int d^dx \, \pi^{ab}(\pounds_v h)_{ab}$, which generate diffeomorphisms of $\Sigma$, and Hamiltonian constraints $H[N]$, given by the complicated expression (\ref{equ:ADMHamilton}) below.   These generate refoliations of the $d+1$ dimensional spacetime.   

Many of the hard questions in quantum gravity and cosmology are tied to the Hamiltonian constraint.  It is then of interest that General Relativity can be reformulated in a way that trades refoliation invariance for a simpler gauge invariance, which is Weyl transformations on the fixed foliation.  To show this, we gauge fix the foliation invariance by imposing the constant mean curvature, $(CMC)$, condition, 
\f
\pi = h_{ab} \pi^{ab} = {\rm \it const}.
\ff
The gauge fixing condition can be expressed as a constraint,
\f
D[\rho] = \int_\Sigma \rho \left ( \pi - \sqrt{h} \frac{\int_\Sigma \pi }{V},
\right )
\ff
where $V= \int_\Sigma \sqrt{h}$ is the volume of $\Sigma$.   Now it is very important to note that $D[\rho]$ generates Weyl transformations of $h_{ab}$ and $\pi^{ab}$ that preserve the overall volume, $V$.  

When General Relativity is gauge-fixed to CMC gauge both $H[N]$ and $D[\rho]$ are imposed.  
We are used to thinking that $H[N]$ is the generator of a gauge symmetry while $D[\rho]$ is merely a gauge fixing condition.  But nothing prevents these roles from being reversed.  
Shape Dynamics arises by interpreting $D[\rho]$ as the generator of a gauge symmetry which is gauge fixed  by $H[N]$. \footnote{New results \cite{Gomes:ADM in CMC} show that  in the phase space of ADM, there are under a number of assumptions only one set of constraints for which one can do this: the set composed of the refoliation constraint and the Weyl constraint.} The physics is the same -- even if we choose a different gauge fixing for $D[\rho]$.  In this way we trade the refoliation invariance generated by $H[N]$ with volume preserving
Weyl transformations generated by $D[\rho]$.  

The key  idea is that Shape Dynamics provides insight into the relationship between gravity and CFT in one less dimension, because it is itself a conformal theory in one less dimension.   This makes the existence of a correspondence between a gravity theory on $\cal M$  and a CFT on its boundary $\Sigma = \partial {\cal M}$ completely transparent because the former theory already enjoys invariance under local scale transformations on $\Sigma$. 

Before getting into the technical implementation of this, we should make one important point.  Note that $D[{\rm \it const}.] $ is identically zero, this ensures $D[\rho]$ only generates volume preserving transformations.  This means that there exists an $N=N_0$ such that $H[N_0]$ is not gauge fixed by $D[\rho]$.  This $H[N_0]$ becomes a global Hamiltonian constraint that generates evolution on the $CMC $ slices.  

\subsection{Shape Dynamics Description of Gravity}

The gauge symmetry trading in gravity is formulated in terms of d-dimensional diffeomorphism constraints ${\rm Diff}[v]=\int d^dx \, \pi^{ab}(\pounds_v h)_{ab}$, which will be unaltered by the gauge symmetry trading mechanism, and the Hamiltonian constraints\footnote{$s$ denotes the signature $+1$ for Euclidean and $-1$ for Lorentzian; and $k$, $\ell$ parametrize the signature and value of the cosmological constant $\Lambda=k\frac{d(d-1)}{2\ell^2}$.}
\begin{equation}\label{equ:ADMHamilton}
 H[N]=\int d^dx N\left(\pi^{ab}\left(h_{ac}h_{bd}-\frac{h_{ab}h_{cd}}{d-1}\right)\frac{\pi^{cd}}{\sqrt{|h|}}+s\left(R[h;x)-\frac{kd(d-1)}{\ell^2}\right)\sqrt{|h|}\right).
\end{equation}
Almost all of these generate on-shell refoliations of spacetime, but, within any given foliation, at least one will generate time evolution within this given foliation. To keep the remaining generator of time evolution, we now trade all but one of the $H[N]$ for d-dimensional Weyl transformations that preserve the total d-volume. For this, we extend phase space by a conformal factor $\phi$ and its canonically conjugate momentum density $\pi_\phi$, and declare this extension pure gauge by imposing that the $\pi_\phi$ be gauge generators. Using the canonical transformation generated by 
\begin{equation}
 F=\int d^dx \left(\Pi^{ab}e^{\frac{4}{d-2}\hat\phi}g_{ab}+\phi\Pi_\phi\right),
\end{equation}
where $\hat \phi:=\phi-\frac{d-2}{2d}\ln\langle e^{\frac{2d}{d-2}\phi}\rangle$ and where $\langle.\rangle$ denotes the mean taken w.r.t. $\sqrt{|h|}$, and using the shorthand $\hat\Omega:=e^{\hat\phi}$ and $\sigma^{ab}=\pi^{ab}-\frac{\pi}{d}h^{ab}$, we find a linking theory with unsmeared Hamiltonian constraints
\begin{equation}
 \begin{array}{c}
   H^\prime(x)\,\,=\\
   \left(\sigma^a_b\sigma^b_a-\frac{\pi^2}{d-1}-\frac{\langle\pi\rangle^2}{d(d-1)}(1-\hat{\Omega}^{\frac{2d}{d-2}})^2|h|+\frac{2\pi\langle\pi\rangle}{d(d-1)}(1-\hat{\Omega}^{\frac{2d}{d-2}})\sqrt{|h|}\right)\frac{\hat{\Omega^{\frac{2d}{2-d}}}}{\sqrt{|h|}}\\
   +s\left(\hat \Omega \left(R+\frac{4(1-d)}{d-2}\Delta\right)\hat\Omega-\frac{kd(d-1)}{\ell^2}\hat{\Omega}^{\frac{2d}{d-2}}\right)\sqrt{|h|}
 \end{array}
\end{equation}
and generators of volume preserving conformal transformations (VPCTs)
\begin{equation}
D[\rho]=\int d^dx\,\rho\,\left(\pi_\phi-\frac{4}{d-2}(\pi-\langle\pi\rangle\sqrt{|h|})\right).
\end{equation}
To verify that we did not change the physical content of the ADM formulation, we impose the gauge fixing condition $\phi\equiv 0$ which allows us to recover the ADM description of gravity immediately.

Shape Dynamics is constructed by imposing the gauge fixing condition $\pi_\phi\equiv 0$, which allows us to simplify the VPCT generators to
\begin{equation}\label{equ:VPCTconstraint}
D[\rho] = \int d^dx\,\rho\,\left(\pi-\langle\pi\rangle\sqrt{|h|}\right).
\end{equation}
The Hamiltonian constraints, which will be eliminated by the phase space reduction, simplify to
\begin{equation}\label{equ:LYE}
 \begin{array}{rcl}
  H^\prime(x)&=&\sigma^a_b\sigma^b_a\frac{\hat{\Omega^{\frac{2d}{2-d}}}}{\sqrt{|h|}}+s\hat \Omega \left(R+\frac{4(1-d)}{d-2}\Delta\right)\hat\Omega\sqrt{|h|}\\
    &&+\left(\frac{\langle\pi\rangle^2}{d(d-1)}+\frac{skd(d-1)}{\ell^2}\right)\hat{\Omega}^{\frac{2d}{d-2}}\sqrt{|h|}.
 \end{array}
\end{equation}
This equation is $\sqrt{|h|}\hat\Omega$ times the Lichnerowicz--York equation \cite{York:york_method_prl} and thus has a unique solution for the variable $\hat\Omega$. Since $\hat \Omega$ is restricted to be volume preserving, we obtain the following defining equations for the single remaining Hamiltonian constraint, which will be the generator of time reparametrizations in the Shape Dynamics description of gravity:
\begin{equation}\label{equ:SDdefi}
  \frac{H^\prime(x)}{\sqrt{|h|(x)}}=H_{SD},\,\,\,\,\,\,\,\,\left\langle\hat\Omega^{\frac{2d}{d-2}}\right\rangle=1,
\end{equation}
where $H_{SD}$ is independent of $x$. Let us conclude this section with the remark that the dictionary between the ADM description and Shape Dynamics is given by ADM in constant mean (extrinsic) curvature (CMC) gauge, which means that $\frac{\pi(x)}{\sqrt{|h|(x)}}={\rm \it const}$. i.e. reduced phase space and the equations of motion of ADM in CMC gauge manifestly coincide with the reduced pahse space and equations of motion of Shape Dynamics in a gauge determined by $\hat \phi_o[h,\pi;x)=1$, where $\phi_o[h,\pi;x)$ denotes the conformal factor that solves the Lichnerowicz--York equation.

\subsection{Special CMC Slices}\label{sec:specialCMC}

The dictionary between the ADM description and Shape Dynamics is given by $ADM$  in CMC gauge. To see this manifest equivalence in the equations of motion, one needs to evolve the ADM system with the CMC lapse $N_{CMC}[h,\pi]$ which is in general a non--local functional of the canonical data $h_{ab},\pi^{ab}$. However, if ADM data satisfies that $R[h;x)$ is a constant then $N={\rm \it const}.$ is a CMC lapse and hence, due to the manifest equivalence with Shape Dynamics, the ADM equations of motion with homogeneous lapse will exhibit VPCT symmetry in this slice. 

To show this, we first notice that the ADM Hamiltonian constraints (\ref{equ:ADMHamilton}) imply that a CMC slice with $R[h;x)=const.$ has constant $\frac{\sigma^a_b\sigma^b_a}{|h|}$. This implies that all coefficients of the Lichnerowicz--York equation (\ref{equ:LYE}) are constants, so homogeneous $\Omega$ is a solution, which implies that $\hat \Omega(x)\equiv 1$. Inserting this into the defining equations (\ref{equ:SDdefi}), we see that $H_{SD}=H(N\equiv 1)$. We can thus use the manifest equivalence  of the ADM equations of motion with Shape Dynamics to predict that the homogeneous lapse evolution of the ADM system in a CMC slice with homogeneous $R[h;x)$ exhibits conformal gauge symmetry. This means, in canonical language, that the VPCT constraint (\ref{equ:VPCTconstraint}) is propagated by the ADM evolution with homogeneous lapse. This can be verified directly evolving by the VPCT constraint (\ref{equ:VPCTconstraint})  with homogeneous lapse. Using the fact that the homogeneous lapse evolution of $\sqrt{
|h|}$ is proportional to $\sqrt{|h|}$ in a CMC slice, we can verify that the VPCT constraint is preserved by checking that $\partial_t \pi$ is a constant multiple of $\sqrt{|h|}$:
\begin{equation}\label{equ:specialCMCeom}
 \begin{array}{rcl}
   \partial_t \pi&=&g_{ab} \partial_t \pi^{ab}+\pi^{ab} \partial_t g_{ab}\\
    &=&  \left(2\pi_{ab}\pi^{ab}-3\pi^2\right)/\sqrt{|h|}-\frac 1 2\left(R\sqrt{|h|}+(\pi_{ab}\pi^{ab}-\frac 1 2 \pi^2)/\sqrt{|h|}\right),
 \end{array}
\end{equation}
where $R$, $\pi\sqrt{|h|}$ and $\pi_{ab}\pi^{ab}/{|h|}$ are homogeneous by assumption. 

Let us conclude this section with a provocative remark. We could have found conformal gauge symmetry of the equations of motion without knowing about Shape Dynamics by directly checking homogeneity of the RHS of \ref{equ:specialCMCeom}. Had we found it this way, we would have wondered where the emergent conformal symmetry came from and we might not have immediately guessed that the underlaying bulk equivalence of the ADM formulation with Shape Dynamics is responsible for the conformal symmetry. In the next section, we will show that the asymptotic conformal symmetry of asymptotic locally AdS spaces is due to the same mechanism: The AlAdS boundary conditions imply that the radial CMC slices have asymptotically vanishing intrinsic Ricci scalar and hence the emergent conformal symmetry at the boundary is simply the gauge symmetry of Shape Dynamics.

\section{Asymptotically Locally AdS and VPCT}\label{sec:AlAdSandVPCT}

The metric of a Euclidean asymptotically locally AdS spacetime takes the asymptotic form (in local coordinates near the conformal boundary $r\to \infty$) 
\begin{equation}\label{reef}
 \begin{array}{rcl}
   ds^2&=&g_{\mu\nu}dx^\mu dx^\nu=dr^2+r^2 \gamma_{ab} dx^adx^b\\
        &=&dr^2+r^2 \gamma^{(0)}_{ab}dx^adx^b+\gamma^{(1)}_{ab}dx^adx^b+\mathcal O(r^{-2}).
 \end{array}
 \label{rdef}
\end{equation}
where we have denoted the \emph{rescaled} spatial metric by $\gamma_{ab}$, and its expansion in radii powers by $\gamma^{(n)}$. The intrinsic metric of $r=const.$ slices is thus $h_{ab}=r^2 \gamma^{(0)}_{ab}+\gamma^{(1)}_{ab}+\mathcal O(r^{-2})$, and in these type of coordinates $g_{ab}=h_{ab}$. 

Moreover, it follows from the fact that alAdS metrics satisfy Einstein's equations that the Riemann tensor takes the asymptotic form 
$$^{(d+1)}R_{\mu\nu\rho\sigma}=-\frac{1}{\ell^2}\left(g_{\mu\rho}g_{\sigma\nu}-g_{\mu\sigma}g_{\rho\nu}\right)+\mathcal O(r^{-2}).$$ 
The intrinsic components in a $r={ \rm \it const}.$ slice are thus 
\begin{equation}
 \begin{array}{rcl}
   ^{(d+1)}R_{abcd}&=&-\frac{1}{\ell^2}\left(g_{ac}g_{db}-g_{ad}g_{bc}\right)+\mathcal O(r^{-2})\\
     &=&-\frac{r^4}{\ell^2}\left(\gamma^{(0)}_{ac}\gamma^{(0)}_{db}-\gamma^{(0)}_{ad}\gamma^{(0)}_{bc}\right)+\mathcal O(1)
 \end{array}
\end{equation}
and the extrinsic curvature of $r=const.$ slices is 
\begin{equation}\label{equ:AlAdSCMC}
 K_{ab}=\frac{r^2}{\ell^2}\gamma^{(0)}_{ab}+\mathcal O(1).
\end{equation}
Inserting this into the Gauss--Codazzi relation for Euclidean signature $^{(d+1)}R_{abcd} = ~^{(d)}\!R_{abcd}-K_{ac}K_{db}+K_{ad}K_{bc}$ we find that the intrinsic Riemann tensor $^{(d)}R_{abcd}=\mathcal O(r^2)$, so the intrinsic Ricci scalar $^{(d)}R=~ h^{ad}h^{bc}\,^{(d)}R_{abcd}$ is
\begin{equation}\label{equ:AlAdSconstR}
  ^{(d)}R=\mathcal O(r^{-2}).
\end{equation}
Equation (\ref{equ:AlAdSCMC}) implies that the $r={\rm \it const}.$ slices are asymptotically CMC and equation (\ref{equ:AlAdSconstR}) implies that these CMC slices have asymptotically homogeneous intrinsic Ricci scalar. We are thus in the special case described in section \ref{sec:specialCMC} and we conclude that the radial evolution of the ADM system becomes manifestly equivalent to the Shape Dynamics evolution at the conformal boundary. The bulk VPCT symmetry of Shape Dynamics thus implies that the radial evolution exhibits VPCT symmetry at the conformal boundary.

Let us conclude this section by looking at the behavior of $I(r)=\int \sqrt{|h|}^{(d)}R[h]$ near the boundary in two and three dimensions: In both cases, $R$ scales as $r^{-2}$ near the conformal boundary, but the volume element scales as $r^2$ in two dimensions, and as $r^3$ in three dimensions. For compact\footnote{We work in Euclidean signature, where the boundary of d+1 AdS is a d-sphere.} $r={\rm \it const}.$ slices follows that $\lim\limits_{r\to\infty}I(r)$ is  finite in two dimensions, but diverges in three dimensions. Similar arguments can be made  for relevant and marginal curvature invariants in higher dimensions, for example for $^{(d)}R^{ab}\,^{(d)}R_{ab}=\mathcal{O}(r^{-4})$. 

\section{Holographic Renormalization and Shape Dynamics}\label{sec:HoloRen}

We will now derive the large volume approximation to the Hamilton--Jacobi function of Shape Dynamics. These results are similar to results obtained in the Hamiltonian approach to holographic renormalization of pure gravity \cite{Skenderis:holoRG_main} and are a generalization of the results obtained in \cite{gryb:gravity_cft} suitable for the context needed here. We conclude this section with a comparison of the two ways to derive these results and explain why the VPCT invariance of Shape Dynamics is compatible with local counter terms and a local conformal anomaly.

\subsection{Classical Shape Dynamics at Large Volume}

In this technical section, we provide some derivations within the Shape Dynamics description of gravity. The starting point are the defining equations (\ref{equ:SDdefi}) for the Shape Dynamics Hamiltonian, which we will derive perturbatively in a large volume expansion. We use this Hamiltonian to find the solution to the Hamilton--Jacobi to the first two orders. The calculations for higher orders require a more sophisticated expansion technique, which goes beyond the scope of this paper.

\subsubsection{Volume Expansion}\label{sec:vol expansion}

It is convenient to isolate the $d$-dimensional volume and its conjugate momentum:
\begin{equation}
 V = \int_\Sigma \sqrt{|h|} \,,\,\,\,\,\,\,   P= \frac 2 d \langle\pi\rangle \,,
\end{equation}
from the other degrees of freedom. For this, we define the fixed-volume metric and its conjugate momentum:
\begin{equation}
   \bar h_{ab} = \left(\frac{V}{V_0} \right)^{-\frac{2}{d}} h_{ab} \,, ~~~   \bar \pi^{ab} = \left(\frac{V}{V_0}\right)^{\frac{2}{d}}\left(\pi^{ab} - \frac 1 d \langle\pi\rangle h^{ab} \sqrt{|h|} \right) \,,
\end{equation}
where $V_0 = \int_\Sigma \sqrt{|\bar h|}$ is some arbitrary but fixed reference volume. The Poisson algebra of the new variables is
\begin{equation}
    \{ V, P \}= 1 \,, ~~~
\left\{\bar{h}_{cd}(x),\bar{\pi }^{ab}(y)\right\}=\frac{1}{2}\delta ^{(c}{}_a\delta ^{d)}{}_b\delta ^{(d)}(x-y)-\frac{1}{d}\frac{\sqrt{\bar h(x)} }{V_0} \,{\bar h}_{cd}(y) \, {\bar h}^{ab}(x) \,.
\end{equation}
 This explicitly isolates the $V$-dependence of the theory. In terms of the new variables,  the defining equations \ref{equ:LYE} and \ref{equ:SDdefi} become\\
\begin{equation}\label{eq:mainSDtransformed}
  \begin{array}{rcl}
    H_\textrm{\tiny SD} &=& {\textstyle  -\left( \frac{d(d-1)sk}{\ell^2} + \frac{d}{4(d-1)} P^2 \right) + \frac{s\left( \bar R - \frac{4(d-1)}{d-2} \hat  \Omega^{-1} \bar \Delta \hat \Omega \right)}{\hat \Omega^{\frac{4}{d-2}}(V/V_0)^{2/d}} 
     +   \frac{ {\bar \sigma}^a_b \,{\bar\sigma}^b_a }{(V/V_0)^2 \Omega^{\frac{4d}{d-2}}|\bar h|} } \,,\\
    \langle \Omega^{\frac{2d}{d-2}}\rangle_0 &=& 1,
  \end{array}
\end{equation}
where barred quantities and spatial averages $\langle \, \cdot \, \rangle_0$ are calculated using $\bar h_{ab}$. 

We will solve  equations  \ref{eq:mainSDtransformed}  by inserting the expansion ansatz
\begin{equation}
     H_\textrm{\tiny SD} = \sum_{n=0}^\infty  {\textstyle \left(\frac{V}{V_0} \right)^{-2n/d} }  H_{(n)} \,, \,\,\,\,\,\,  \hat\Omega^{\frac{2d}{d-2}} = \sum_{n=0}^\infty   {\textstyle \ \left(\frac{V}{V_0} \right)^{-2n/d}  } \omega_{(n)} \,,
\end{equation}
and solving order by order in $V^{-2/d}$. Using this expansion, the second line of \ref{eq:mainSDtransformed} is trivially solved by
\begin{equation}
    \langle\omega_{(0)}\rangle_0 = 1  \,,\,\,\,\,\,    \langle\omega_{(n)}\rangle_0 = 0 \,, ~ \forall ~ n \geq 1 \,.\label{eq:mean omega}
\end{equation}
We can now outline the procedure for finding the solution order-by-order:
\begin{itemize}
\item {\bf For $\mathbf {n= 0}$}, we have trivially
\begin{equation}
    H_{(0)} = - \left( \frac{d(d-1) \,  s \, k}{\ell^2} + \frac d {4(d-1)} P^2 \right) \,.
\end{equation}

\item {\bf For $\mathbf {n=1}$}, we observe that the equations imply that the conformal factor is chosen such that $R$ is homogeneous. This is known as the Yamabe problem, which has a solution on compact manifolds without boundary \cite{YamabeProblem}. The equations thus fix a conformal gauge (Yamabe gauge) such that
\begin{equation}
    R(\tilde h) = \textrm{\it const.} \equiv \tilde R \,.
\end{equation}
We indicate this section in the conformal bundle using a tilde, \emph{e.g.} $\tilde h_{ab}$. This leads to
\begin{equation}
    H_{(1)} =  s \, \tilde R \,, \,\,\,\,\,\, \omega_{(0)} = 1 \,. 
\end{equation}

\item {\bf For $\mathbf {n=2}$}, we use the expansion
\begin{equation}
    \hat\Omega^{\frac{2d}{d-2}} = 1 + \left( \textstyle \frac{V_0}V \right)  \omega_{(1)} + ... \,,
\end{equation}
and get
\begin{equation}\label{equ:hn2prelim}
    H_{(2)} = -\frac{2s}{d} \left( \tilde R + (d-1) \tilde \Delta \right)  \omega_{(1)} \, .
\end{equation}
Taking the mean and using integration by parts to drop boundary terms ($\Sigma$ is compact without boundary) we get
\begin{equation}
    H_{(2)} = -\frac {2 s\tilde R} d \langle\omega{(1)}\rangle_0 = 0 \,.
\end{equation}
Inserting this into \ref{equ:hn2prelim} gives
\begin{equation}
    \left( \tilde R + (d-1) \tilde \Delta \right) \omega_{(1)} = 0 \,.
\end{equation}
This equation admits the solution
\begin{equation}
    \omega_{(1)} = 0\,.
\end{equation}
For negative Yamabe class, this solution is not unique if $\tilde R$ happens to be in the (discrete) spectrum of $\tilde \Delta$. 

\item {\bf For $\mathbf{n= 3 \,\,...\,\, (d-1)}$}, the same reasoning will apply. Using the result $\omega_{(n-2)} = 0$, we can now use the expansion
\begin{equation}
    \hat\Omega^{\frac{2d}{d-2}} = 1 + \left( \textstyle \frac{V_0}V \right)^n  \omega_{(n)} + ... \,,
\end{equation}
which leads to
\begin{equation}
    H_{(n)} = -\frac{2s}{d} \left( \tilde R + (d-1) \tilde\Delta \right)  \omega_{n-1} \,.
\end{equation}
Taking the mean leads to:
\begin{equation}
 H_{(n)} = 0 \,,
\end{equation}
so that
\begin{equation}
    \left( \tilde R + (d-1) \tilde \Delta \right) \omega_{(n-1)} = 0 \,,
\end{equation}
which, again,  has the solution
\begin{equation}
    \omega_{(n-1)} = 0 \,.
\end{equation}

\item {\bf For $\mathbf{n = d}$},  the solution can still easily be worked out using the previous expansions for $n = d$ and including the ${{\tilde \sigma}^a}_b \, {{\tilde \sigma}^b}_a$ term. The resulting equation is
\begin{equation}
    H_{(d)} = -\frac{2s}{d} \left( \tilde R + (d-1) \tilde \Delta \right)  \omega_{(d-1)} + \frac{{{\tilde \sigma}^a}_b \, {{\tilde \sigma}^b}_a}{|\tilde h|} \,.
\end{equation}
Taking the mean, we get
\begin{equation}
    H_{(d)} = \left\langle\frac{{\tilde \sigma}^a_b \, {\tilde \sigma}^b_a}{|\tilde h|}\right\rangle \,.
\end{equation}
$\omega_{(d-1)}$ can then be solved by inverting the following equation
\begin{equation}
   \Delta_c \omega_{(d-1)}  -\frac d {2s(d-1)}\frac{{\tilde \sigma}^a_b \, {\tilde \sigma}^b_a}{|\tilde h|} = \left\langle\Delta_c \omega_{(d-1)}  -\frac d {2s(d-1)}\frac{{\tilde \sigma}^a_b \, {\tilde \sigma}^b_a}{|\tilde h|}\right\rangle \,,
\end{equation}
where $\Delta_c$ is the $d$-dimensional conformal Laplacian
\begin{equation}
    \Delta_c = \tilde \Delta + \frac{\tilde R}{(d-1)} \,.
\end{equation}
Thus, all higher order terms will be non--local because they will involve inverting the conformal Laplacian.
\end{itemize}
Collecting the first three non-zero terms, we get
\begin{equation}\label{eq:hg v exp}
   H_\textrm{\tiny SD} = -\left( \textstyle \frac{d(d-1)sk}{l^2} + \frac d {4(d-1)} P^2  \right) + s \tilde R \left( \textstyle \frac{V_0}{V}\right)^{2/d} + \left\langle\frac{ {\tilde \sigma}^a_b \, {\tilde \sigma}^b_a}{|\bar h|}\right\rangle\left(\textstyle \frac{V_0}{V}\right)^2 + \mathcal O\left( \textstyle  \frac{V}{V_0} \right)^{\frac 2 d -4}  \,.
\end{equation}

\subsubsection{Hamilton--Jacobi Equation}\label{HJ eqn}

Using the substitutions
\begin{equation}
   P \to \frac{\delta S}{\delta V}
\end{equation}
and the fact that, for a VPCT--invariant $S$, one can use\footnote{Notice that the second substitution is by no means trivial: it involves the calculation
$\bar \pi^{ab}  = \left(V/V_0\right)^{2/d} \left( \frac{\delta S }{\delta h_{ab}} - \frac P 2 \sqrt h \, h^{ab} \right) = \frac{\delta S }{\delta \bar h_{ab}} $, using the fact that $\frac{\delta S }{\delta h_{ab}} = \left( V/V_0 \right)^{-2/d} \frac{\delta S }{\delta \bar h_{ab}}  + \frac{\delta S }{\delta V} \frac{\delta V } { \delta h_{ab}}$ and $ \frac{\delta V } { \delta h_{ab}} = \frac 1 2 \sqrt h \; h^{ab}$. Furthermore, it requires the realization that for a VPCT--invariant $S$, one can ignore the variation of the Yamabe conformal factor.}
\begin{equation}\label{eq:hj sub}
   \tilde \sigma^{ab} \to \frac{\delta S}{\delta \tilde h_{ab}} - \frac 1 d \left\langle \tilde h_{ab} \frac{\delta S}{\delta \tilde h_{ab}}\right\rangle \tilde h^{ab} \sqrt{|\tilde h|},
\end{equation}
where $S= S(h_{ab}, \alpha^{ab})$ is the HJ functional, depending on the metric $h_{ab}$ and on $d(d+1)/2$ integration constants $\alpha^{ab}$, we can solve the Hamilton--Jacobi equation associated to Eq. (\ref{eq:hg v exp})
\begin{equation}
    0 =-\left( \frac{d(d-1)sk}{\ell^2} + \frac d {4(d-1)} \left(\frac{\delta S}{\delta V} \right)^2  \right) + s \tilde R \left(\frac{V_0}{V}\right)^{2/d} + \left\langle\frac{\delta S}{\delta \tilde h_{ab}}\frac{\delta S}{\delta \tilde  h^{ab}}\right\rangle\left(\frac{V_0}{V}\right)^2 + ...
\end{equation}
order by order in $V$ using the ansatz
\begin{equation}
    S = \sum_{n=0}^\infty \left(\textstyle \frac{V}{V_0}\right)^{1 - \frac{2n} d} S_{(n)} \,,
\end{equation}
for odd $d$ or
\begin{equation}
 S = \log\left( \textstyle \frac V {V_0} \right) S_{(\frac d 2)} + \sum_{n\neq \frac d 2}^\infty \left( \textstyle \frac{V}{V_0}\right)^{1 - \frac{2n} d} S_{(n)} \,,
\end{equation}
for even $d$ (because the previous ansatz is not valid with even $d$).

We can get a recursion relation for the solution by taking the asymptotic boundary condition
\begin{equation}
    \lim_{(V/V_0) \to \infty} S = S_{(0)} = \textrm{\it const.}
\end{equation}
Using this we get:
\begin{itemize}
\item {\bf For $\mathbf{n=0}$}, the solution is trivial
\begin{equation}
    S_{(0)} = \pm 2 \sqrt{-sk} \frac {(d-1)} \ell \,.
\end{equation}
\item {\bf For $\mathbf{n=1}$}, the solution is equally straightforward. The result for $d\neq 2$ is
\begin{equation}
    S_{(1)} = \pm \frac{\ell \, s \,\tilde R}{(d-2)\sqrt{-sk} } \,.
\end{equation}
In $d=2$ this term gives the conformal anomaly. It is found to be
\begin{equation}
    S_{(1)}^{d=2} = \pm \frac{\ell \, s \,\tilde R}{2\sqrt{-sk}} \,.
\end{equation}

\item {\bf For $\mathbf{n=2}$},  we use
\begin{equation}
    \left\langle\frac{\delta S_{(1)}}{\delta \tilde h_{ab}} \frac{\delta S_{(1)}}{\delta\tilde h^{ab}} \right\rangle = -\frac {s \, \ell^2}{k(d-2)^2} \left( \left\langle\tilde R^{ab} \tilde R_{ab}\right\rangle - \frac {\tilde R^2} d \right) \,,
\end{equation}
and find
\begin{equation}
    S_{(2)} = \mp \frac s {\sqrt{-sk}} \frac{\ell^3}{(d-4)(d-2)^2} \left[ \left\langle\tilde R^{ab} \tilde R_{ab}\right\rangle - \frac d {4(d-1)} \tilde R^2 \right] \,.
\end{equation}
In $d=4$ this gives the anomaly. It is
\begin{equation}
    S_{(2)}^{d=4} = \mp \frac s {\sqrt{-s \, k}} \frac{\ell^3}{8} \left( \left\langle\tilde R^{ab} \tilde R_{ab}\right\rangle - \frac 1 3 \tilde R^2 \right) \,.
\end{equation}

\item {\bf For $\mathbf{n>2}$},
If we ignore higher order terms in the $V$-expansion of $H_\textrm{\tiny SD}$ then we get a compact recursion relation for $S_{(n)}$
\begin{equation}
    S_{(n)} = \pm \frac \ell {(d-2n)\sqrt{-sk}} \sum^{p+q = n}_{p,q > 0} \left[ \left\langle\frac{\delta S_{(p)}} { \delta \tilde h_{ab}}  \frac{\delta S_{(q)}} {\delta \tilde h^{ab}}\right\rangle - \frac{(d-2p)(d-2q)}{4(d-1)d} S_{(p)}S_{(q)} \right]+...\,.
\end{equation}
In general though, there will be contributions from higher order terms that depend on the inverse of $\Delta$ but these are not important for $d<5$.
\end{itemize}

\subsection{Comparison with Holographic Renormalization Results}

The Hamiltonian approach to holographic renormalization \cite{Skenderis:holoRG_main} uses the AdS/CFT correspondence to relate the near boundary behavior of the classical Hamilton--Jacobi functional of a gravity theory to the partition function of a CFT in the strong coupling limit. In particular, the radial evolution near the conformal boundary of the Euclidean asymptotically locally AdS Hamilton--Jacobi functional, $S$, is related to the renormalization of a dual CFT on the boundary: in the large volume limit $V\to\infty$, the divergent part of $S$ is identified with counter terms while a standard argument shows that the logarithmic term can be identified with the integral of the conformal anomaly.\footnote{A slightly different procedure allows for the calculation of the local form of the anomaly \cite{Skenderis:holo_RG}, which is consistent, in the near boundary limit, to the results presented here for the reasons described below.} These terms are local in the sense that the counter term Lagrangian 
contains local curvature invariants. 

The results of the previous subsection show that the large volume limit of the Shape Dynamics Hamilton--Jacobi functional takes the same form as the results from General Relativity described in \cite{Skenderis:holoRG_main} if the conformal factor of the intrinsic metric $h_{ab}$ is replaced with the Yamabe conformal factor. This choice of conformal frame has three important consequences for the Shape Dynamics Hamilton--Jacobi functional:

\begin{enumerate}
 \item It ensures that the Shape Dynamics Hamilton--Jacobi functional is manifestly VPCT-invariant, i.e. invariant under d-dimensional Weyl transformations that preserve the total d-volume.
 
 \item In the large volume limit, the divergent terms, which are integrals of \emph{local} terms for Yamabe metrics, are turned into non-local terms for metrics that are not Yamabe.
 
 \item{} The integrated form of the anomaly is manifestly VPCT in its local form and hence has the same form in any conformal gauge.
\end{enumerate}

We have to address the subtle issue we mentioned in the introduction regarding the difference between the local anomaly, (\ref{local}) 
that appears in a CFT and the integrated form of the anomaly that arises in Shape Dynamics. The bulk-bulk equivalence between the ADM and the Shape Dynamics description of General Relativity raises the question of how this discrepancy between the ADM and Shape Dynamics descriptions is possible?

The answer to this question lies in the following: despite the fact that all observables of Shape Dynamics coincide with observables of the ADM description, the Shape Dynamics equations of motion coincide with the ADM equations of motion only in the dictionary, i.e. only when the ADM system is in a CMC slice and evolved with the CMC lapse. We saw in section \ref{sec:AlAdSandVPCT} that the radial evolution of the ADM system satisfied these conditions only at the conformal boundary, but that these conditions are violated at any distance from the boundary. This provides us with a nontrivial consistency check: the integrals of the non-local VPCT-invariant terms derived in the large volume expansion of Shape Dynamics have to coincide with the integrals of the local terms derived from the radial ADM evolution at the boundary. 


The explicit evaluation of the curvature invariants at the boundary (performed at the end of section \ref{sec:AlAdSandVPCT}) confirm this assertion. For example, in $d=2$, the VPCT invariant Shape Dynamics anomaly falls-off as: 
$$ \frac{1}{V}\int \sqrt{|h|}^{(d)}R[h]=\mathcal{O}(r^{-2}) $$ 
which is the same fall-off rate for the local anomaly when expressed in terms of the induced boundary metric $G_{ab}$ (rather than the rescaled boundary metric $\gamma^{(0)}_{ab}$ used in standard holographic renormalization). This ensures that the integrals coincide \emph{at the boundary}. 
The same mechanism works for the higher dimensional counter-terms, again using $G_{ab}$, we find for example 
$$ \left( \left\langle\tilde R^{ab} \tilde R_{ab}\right\rangle - \frac 1 3 \tilde R^2 \right)=\mathcal{O}(r^{-4})= R^{ab} R_{ab} - \frac 1 3  R^2 $$ 
 
This shows how the VPCT invariance of Shape dynamics is compatible with local counter terms in the radial ADM evolution: The equations of motion for $G_{ab}$ of General Relativity and Shape Dynamics coincide manifestly when the boundary is approached. Since the boundary metric $G_{ab}$ gives only access to the integrals of the curvature invariants, one might think that Shape Dynamics provides less information than standard holographic renormalization, which is expressed in terms of the rescaled metric $\gamma^{(0)}_{ab}$, which provides access to local curvature invariants. However, Shape Dynamics provides actually more, once one moves away form the boundary. It states that if we replaced radial evolution into the bulk with CMC evolution into the bulk, then we would not only find VPCT invariance at the boundary, but also in the bulk.

We can make two further remarks on the results found so far.
\begin{itemize}
\item{} It should be emphasized that in the case that the bulk manifold is pure Anti-de~Sitter spacetime (rather then AlAdS), the metric and curvatures are homogeneous and the constant $r$ slices are CMC in the bulk.  In this case, the Hamilton--Jacobi functions of Shape Dynamics coincides with that of General Relativity exactly in the bulk as well as the boundary.   
\item{}Our results indicate the emergence of full Weyl invariance on the boundary as the volume dependence disappears when the limit 
$V \rightarrow \infty$ is taken.  That is, after removal of the divergent terms, the remaining finite terms in $S$ are fully Weyl-invariant in the limit. However, it is known that a classical field theory defined on a manifold with an arbitrary metric, $h_{ab}$, that is Weyl invariant will have global conformal symmetry under $SO(2,d)$ when that metric is taken to be flat \cite{Faci:weyltoconformal}.\footnote{This can be extended to a conformally flat metric (as opposed to flat), by using the complete set of  conformal Killing vector fields in place of the coordinates in flat space. } This will apply to the finite parts of $S$ in the $V \rightarrow \infty$ limit.  That function then has several properties needed to describe the semiclassical limit of a CFT.
\end{itemize}

\section{Remarks on Wilsonian Renormalization}\label{sec:Speculation}

So far, we used the classical bulk equivalence of General Relativity and Shape Dynamics to explain why classical gravity is related to a classical conformal filed theory near the conformal boundary of asymptotically locally AdS spaces. This is however not  how the AdS/CFT correspondence is used in practice, where the relation between the radial evolution of classical gravity and the RG flow of a CFT in the strong coupling limit is used. We will now present a heuristic observation that can be used to turn the classical correspondence we have so far described into a quantum correspondence. The purpose of this section is thus to turn the AdS/CFT logic upside down, as it was suggested in \cite{Koslowski:SD_EFT}. Whereas the usual logic assumes the AdS/CFT conjecture and derives the holographic renormalization group equations in a semiclassical limit, we go the other way: We use the proven classical equivalence between General Relativity and Shape Dynamics and explore the consequence of this equivalence for an 
assumed gravity 
partition function with a recognizable semiclassical limit near the conformal boundary. 

We start with assuming the existence of a Shape Dynamics boundary amplitude $Z_V[\bar h]$, where $\bar h_{ab}$ denotes the rescaled metric at CMC volume $V$. This boundary amplitude is supposed to be obtained as the solution to the semiclassical defining equations of Shape Dynamics in a Euclidean asymptotically locally AdS space in the limit $V\to \infty$. We suppose that the semiclassical limit is recognizable in the sense that the classical Shape Dynamics constraints are turned into operators acting on $Z_V[\bar h]$ through the replacement $\pi^{ab}(x) \to i \hbar \frac{\delta}{\delta h_{ab}(x)}$. Since the boundary conditions imply that the classical Shape Dynamics Hamiltonian constraint asymptotes into the homogeneous lapse Hamiltonian in the limit $V\to\infty$, we impose in the large $V$ limit the radial Wheeler--DeWitt equation:
\begin{equation}\label{equ:radialWdW}
  \int d^dx \left(-\hbar^2(h_{ac}h_{bd}-\frac{1}{d-1}h_{ab}h_{cd})/\sqrt{|h|}\frac{\delta^2}{\delta h_{ab}\delta h_{cd}}+...\right)Z_V[\bar h]=0+\mathcal O(\hbar).
\end{equation}
Moreover, the classical VPCT constraints of Shape Dynamics lead to
\begin{equation}\label{equ:conformalWard}
 h_{ab}(x)\frac{\delta}{\delta \bar h_{ab}(x)}Z_V[\bar h]=0+\mathcal O(\hbar).
\end{equation}
Equation (\ref{equ:conformalWard}) has an important consequence: it states that the wrong sign in the kinetic term of equation  (\ref{equ:radialWdW}) 
is $\mathcal O(\hbar)$, except for the derivative w.r.t. $V$. We can thus rewrite equation (\ref{equ:radialWdW}) as a second order evolution equation in $V$ with a positive definite kinetic term $\int d^dx \hbar^2\,h_{ac}h_{bd}/\sqrt{|h|}\frac{\delta^2 Z_V[\bar h]}{\delta h_{ab}\delta h_{cd}}$. This term looks like the Schwinger--Dyson equation for a mass term in $d$ dimensions, which is the UV--limit of an IR suppression term as it is used in the exact renormalization group framework, see e.g. \cite{Wetterich:exact_evolution}. If we interpret the extra terms $...$ in equation (\ref{equ:radialWdW}) as the remnant of a particular scheme, then we have an argument that relates the radial evolution with renormalization group flow\footnote{Notice the important difference that usual renormalization group equations are first order equations, while we derive a second order equation in 
$V$. In a semiclassical regime, one can argue that $Z_V$ will be a sum of an ``incoming'' and an ``outgoing'' wave function in $V$. The interpretation of our second order equation as a renormalization group equation requires that we restrict ourselves to the ``outgoing'' summand. This was also discussed by
Freidel in \cite{Freidel}.}. We want to warn the reader that this heuristic argument is at best a plausibility argument: the validity of our assumptions have to be checked before the argument worked out in a particular application. However, we want to point out that the VPCT--invariance of Shape Dynamics is the {\bf essential} ingredient that allows us to reinterpret radial evolution of semiclassical gravity in asymptotically locally AdS d+1 as the UV--limit of renormalization group flow of a classical CFT in d dimensions. 

If we interpret the correspondence between radial evolution of Shape Dynamics and Wilsonian renormalization literally, then we might be able to reconcile the nonlocal VPCT-invariant counter terms that found in the present paper with the usual local counter terms. This is due to the fact that the fixed point of a Wilsonian flow equation may differ from the critical bare action by a scheme dependent one-loop determinant (see e.g. \cite{Reuter:1996eg}). The reconciliation would thus follow from the conjecture that the radial evolution of Shape Dynamics is equivalent to a particular Wilsonian scheme.

\subsection{Generic Recipe}\label{sec:recipe}

The combination of the heuristic quantum argument presented in this section with the classical symmetry trading provides a generic mechanism that can be used to construct dualities of the type of holographic renormalization. This recipe can be summarized as follows:
\begin{enumerate}
 \item Use the classical linking theory formalism to derive a bulk--bulk equivalence between two classical gauge theories.
 \item Construct the dictionary between these two classical gauge theories.
 \item Assume a partition function with a recognizable semiclassical limit in dictionary gauge. Then use the classical dictionary to reinterpret the semiclassical quantum equations.
\end{enumerate}
We believe in the value of this generic recipe, in part because the calculations in section~\ref{sec:HoloRen} show that this generic recipe can be used to find a number of results that are usually attributed to the AdS/CFT correspondence.

\section{Conclusions}\label{sec:Conclusions}

In this paper we have shown that the Hamilton-Jacobi function of Shape Dynamics has, when evaluated on the boundary of an $AlAdS$ spacetime,
with divergent terms removed, several properties needed to posit a correspondence to the semiclassical limit  of the effective action of a CFT.  By invoking gauge symmetry trading this explains very generally why the  foliation invariance of General Relativity manifests itself as conformal invariance on the boundary of an $AlAdS$ spacetime.  Thus, gauge symmetry trading provides a  deep and general explanation of why there exist correspondences between gravitational theories invariant under spacetime diffeomorphisms and conformal field theories in one lower dimension.   As a check on the general argument we also confirmed that Shape Dynamics reproduces the precise forms and coefficients found for the integrals of trace anomalies using the methods of holographic renormalization \cite{Skenderis:holo_RG,Skenderis:holoRG_main}.

Some concluding remarks are in order:

\begin{itemize}

\item We provide a mathematical mechanism (trading of gauge symmetries) as a construction principle for classical dualities. This mechanism provides a complete one--to--one dictionary between the physical predictions of two at first sight very different looking gauge theories. Moreover, the dictionary proves that all local physical predictions of these two classical gauge theories coincide. 
The symmetry trading mechanism has previously been used to show that spacetime General Relativity is physically equivalent to Shape Dynamics by trading refoliation invariance of spacetime for local spatial conformal invariance.

\item In this paper we showed that the classical bulk--bulk equivalence of Shape Dynamics and General Relativity explains some aspects of classical AdS/CFT, in particular conformal symmetry of the boundary theory and the form of the classical Hamilton--Jacobi functional. However, Shape Dynamics does not explain specific correspondences between particular CFT's and their dual gravitational theories.  We note that our results reproduce, but do not predict, 
the correspondence between pure General Relativity and ${\cal N} =4$ super-Yang--Mills theory in the $N\to \infty$ limit which was found by holographic renormalization group methods \cite{Skenderis:holo_RG,Skenderis:holoRG_main}.  

\item A possible extension of the work in this paper is to explore the conjecture that the correspondence we have demonstrated here extends to a stronger correspondence between a quantization of Shape Dynamics and a quantum CFT. The evidence we possess for an extension of the correspondence into the quantum regime is the apparent absence of anomalies (for odd dimensions) of our spatial conformal transformations  and the expected effect of matter fields (see appendix \ref{app:matterfields}). In spite of their classical correspondence,  the quantization of Shape Dynamics  is unlikely to coincide with a quantization of General Relativity due to the very different structure of their constraints.

\item The gauge symmetry trading mechanism is very generic. We thus expect that it can be successfully applied to attack a variety of problems that are not part of AdS/CFT. A first example of this sort was explored in \cite{Gomes:FAQs}, where the $U(1)$--gauge symmetry of classical electromagnetism was traded for shift symmetry. A main motivation for this paper was to advertises the power of the symmetry trading mechanism to researchers interested in dualities.

\end{itemize}

\section*{Acknowledgements}

We are grateful to Julian Barbour for encouragement, to Kostas Skenderis for comments on a draft of this manuscript and to
Paul McFadden for helpful conversations. We also thank Jacques Distler and Laurent Freidel for pointing out an oversight in the first draft of this paper. S.G.'s work was supported by the Natural Science and Engineering Research Council (NSERC) of Canada and by the Netherlands Organisation for Scientific Research (NWO) (Project No. 620.01.784). Research at Perimeter Institute is supported by the Government of Research at Perimeter Institute is supported by the Government of Canada through Industry Canada and by the Province of Ontario through the Ministry of Economic Development and Innovation.  This research was also partly supported by grants from NSERC, FQXi and the John Templeton Foundation.

\newpage

\begin{appendix}

\section{APPENDIX: General Mechanism for Gauge Symmetry Trading}\label{app:sym_trading}

\subsection{Symmetry Trading and Linking Gauge Theories}

A (special) linking gauge theory is a canonical gauge theory $(\Gamma,\{.,.\},H,\{\chi_\mu\}_{\mu\in\mathcal M})$, where the phase space and Poisson structure $(\Gamma,\{.,.\})$ can be written as a direct product of two mutually commuting phase spaces $(\Gamma_o,\{.,.\}_o)$ and $(\Gamma_e,\{.,.\})$ and where the first class (coisotropic) constraint surface $\mathcal C=\{x\in\Gamma:\chi_\mu(x)=0,\mu\in\mathcal M\}$ can be specified by three disjoint sets of irreducible constraints
\begin{equation}
 \begin{array}{rcl}
   \chi^1_\alpha&=&\phi_\alpha-\sigma_\alpha(p,q)\\
   \chi_2^\alpha&=&\rho^\alpha(p,q)-\pi^\alpha\\
   \chi^3_\nu&=&\chi^3_\nu(p,q)+more,
 \end{array}
\end{equation}
where $more$ vanishes when either $\phi_\alpha\equiv 0$ or $\pi^\alpha\equiv 0$ holds. $(p,q)$ denote here local Darboux coordinates for $(\Gamma_o,\{.,.\}_o)$, while $\phi_\alpha$ denote local position coordinates on $(\Gamma_e,\{.,.\}_e)$, whose canonically conjugate momenta are $\pi^\alpha$. A canonical gauge theory of this kind can be gauge--fixed in two very interesting ways (1) by imposing $\phi_\alpha\equiv 0$ and a partial gauge fixing condition for the constraints $\chi_2$ and (2) by imposing $\pi^\alpha\equiv 0$ as a partial gauge fixing condition for the constraints $\chi_1$. The two partial gauge fixings lead to the same reduced phase space $\Gamma_o$ and the Dirac bracket associated with this phase space reduction reduces to $\{.,.\}_o$, i.e. the Poisson bracket on $\Gamma_o$. The two theories have however distinct constraints and Hamiltonians.

The partial gauge fixing (1) reduces the constraints to $\sigma_\alpha(p,q)$ and $\chi^3_\mu(p,q)$ and the Hamiltonian is $H(p,q,\sigma_\alpha(p,q),0)\approx H(p,q,0,0)$. The partial gauge fixing (2) on the other hand reduces the constraints to $\rho^\alpha(p,q)$ and $\chi^3_\mu(p,q)$ and the Hamiltonian is $H(p,q,0,\rho^\alpha(p,q))\approx H(p,q,0,0)$. The two gauge theories describe the same physics, since each of them is obtained as a partial gauge fixing of the same linking theory, but the gauge symmetries of the two theories differ. In other words: the gauge generators $\sigma_\alpha(p,q)$ of (1) can be traded for the gauge generators $\rho^\alpha(p,q)$ without changing any physical prediction of the theory.

\subsection{Dictionary and Observable Equivalence}

The manifest equivalence of the two descriptions can be seen by further gauge fixing: imposing $\rho^\alpha(p,q)\equiv 0$ as a gauge fixing condition on (1) and working out the phase space reduction gives precisely the same reduced phase space $\Gamma_{red}$ and Dirac bracket $\{.,.\}_D$ as it is obtained by imposing $\sigma_\alpha(p,q)\equiv 0$ on (2). This reduced gauge theory $(\Gamma_{red},\{.,.\}_D,H_{red},\{\chi^3_\nu\}_{\nu \in \mathcal N})$ serves as a dictionary between the two gauge theories, where the two descriptions manifestly coincide. 

A constructive procedure to construct the dictionary goes as follows: Consider the Poisson algebra of observables of the linking theory. We can then use the fact that the phase space reduction is a Poisson--isomorphism for observables\footnote{An observable is an equivalence class of gauge--invariant phase space functions, where two phase space functions are called equivalent, if they coincide on the constraint surface. For simplicity, we will slightly abuse notation and write representative phase space functions $O$ rather than equivalence classes $[O]_\sim$.} to calculate entries of the dictionary one--by--one: Pick an observable $O$ of the linking theory and insert the two phase space reductions. Thus every linking theory observable $O$ relates an observable of (1) with an observable of (2) by
\begin{equation}
 O^{(1)}:=\left.O\right|_{(\phi\equiv 0,\pi\equiv\pi_o(p,q))}\,\leftrightarrow\,O^{(2)}\left.O\right|_{(\phi\equiv\phi_o(p,q),\pi\equiv 0)},
\end{equation}
where the $O^{(i)}$ denote an observable of the system (i). Notice that this procedure is surjective and a Poisson--isomorphism for observables, i.e. if one starts with a complete observable algebra of the linking theory, one obtains a complete dictionary, which is given by a Poisson--isomorphism between the observable algebras of description (1) and description (2).

\subsection{Construction of Linking Gauge Theories}

A useful construction principle for linking gauge theories is Kretschmannization (also called the general St\"uckelberg mechanism), which is most simply explained for the action of an Abelian group $\mathbb G$ on configuration space. Let us denote local coordinates on configuration space by $q_i$ and consider an Abelian group action $g(\phi): q_i \mapsto Q_i(q,\phi)$ that is locally parametrized by group parameters $\phi_\alpha$. We can implement gauge invariance under this group action in any gauge theory $(\Gamma_o,\{.,.\}_o,H_o,\{\chi_\mu\}_{\mu\in\mathcal M})$ using the following steps:
\begin{enumerate}
 \item Extend phase space with the cotangent bundle of the group, which we will denote by $(\Gamma_e,\{.,.\})$, and declare the extension to be pure gauge by introducing the auxiliary first class constraints $\pi^\alpha\approx 0$, where the $\pi^\alpha$ denote the momenta canonically conjugate to the group parameters $\phi_\alpha$.
 \item Perform the ``Kretschmannization'' canonical transformation generated by $F=Q_i(q,\phi)P^i+\phi_\alpha\Pi^\alpha$, where capital letters denote the transformed variables. This canonical transformation changes the form of the Hamiltonian $H_o(p,q) \to H(p,q,\phi)$ and the gauge generators $\chi_\mu(p,q)\to\tilde\chi_\mu(p,q,\phi)$, but most importantly, it implies that the auxiliary gauge generators transform as
 \begin{equation}
   \pi^\alpha \to \pi^\alpha - p^i W^\alpha_i(q),
 \end{equation}
\end{enumerate}
where the $W^\alpha_i(q) \partial^i$ are the vector fields that generate the $\mathbb G$ action on configuration space. If now a subset $\{\tilde \chi_\alpha(p,q,\phi)\}_{\alpha\in\mathcal A}$ of the transformed gauge generators $\tilde \chi_\mu(p,q,\phi)$ can be uniquely solved for the $\phi_\alpha$, then we have a linking gauge theory. To see which theories are linked let us, without loss of generality, assume that the group parametrization is such that $\phi_\alpha\equiv 0$ denotes the unit element of $\mathbb G$, so $Q_i(q,0)=q_i$. Then imposing the gauge fixing condition $\phi_\alpha\equiv 0$ and working out the phase space reduction reduces the theory to the system $(\Gamma_o,\{.,.\}_o,H_o,\{\chi_\mu\}_{\mu\in\mathcal M})$ we started with. On the other hand, imposing the gauge fixing condition $\pi^\alpha\equiv 0$ yields a system in which the gauge generators $\tilde \chi_\alpha(p,q)$ have been traded for the gauge generators $p^i W^\alpha_i(q)$, which implement the $\mathbb G$ action we have put into 
our construction. 

We thus have a very generic construction principle: We can start with an arbitrary gauge theory and the mechanism allows us to trade a subset of its gauge generators $\chi_\alpha(p,q)$ for a different set of gauge symmetry generators without changing the physical description. The only nontrivial requirement is that the Kretschmannized gauge generators $\tilde\chi_\alpha(p,q,\phi)$ can be uniquely solved for the $\phi_\alpha$.

\section{APPENDIX: Scalar Field ($d\leq 4$)}\label{app:matterfields}

The goal of this appendix is to illustrate how, in the presence of matter fields, the coefficients in the volume expansion of Hamilton's principle function, $S[h_{ab}]$, for Shape Dynamics will depend upon the matter fields. We will demonstrate this using the simple example of a single real scalar field, $\varphi$.

The matter Hamiltonian for this scalar field is:
\begin{equation}
    H_\textrm{m} (\varphi,h_{ab}) = \frac {\pi_\varphi^2}{\sqrt{|h|}} + \left( U(\varphi) - s h^{ab} \nabla_a \varphi \nabla_b \varphi \right) \sqrt {|h|}.
\end{equation}
Adding this to the Hamiltonian constraint and performing the phase space extension and canonical transformation described in the main text, we obtain the Shape Dynamics Hamiltonian coupled to a scalar field
\begin{equation}
  \begin{array}{rcl}
  H_\textrm{SD} &=& -\left( \frac{d(d-1)sk}{l^2} + \frac{d}{4(d-1)}P^2 + U(\varphi) \right) \\
   && + s \; \hat \Omega^{- \frac{4}{d-2}} \left( \tilde R - \bar h^{ab} \bar \nabla_a \varphi \bar \nabla_b \varphi - \frac{4(d-1)}{d-2} \hat \Omega^{-1} \nabla^2_{\bar g} \hat\Omega \right) \left(\frac{V_0} V  \right)^{2/d}  \label{HsdScalar}\\
   &&    + \hat \Omega^{- \frac{4 d }{d-2}} \frac 1 {\bar g} \left( {\bar \sigma}^a_b \, {\bar\sigma}^b_a + \pi_\varphi^2\right) \left(\frac{V_0} V  \right)^{2}.
   \end{array}
\end{equation}

In \cite{Gomes:MatterPaper}, it was shown that a consistent coupling of matter to Shape Dynamics requires that the matter fields be invariant under VPCT. They can, however, transform non-trivially under homogeneous conformal transformation, and the weight of this transformation represents the anomalous scaling, $\Delta$, of the matter fields. For a real scalar field, we find that the dilatation operator has the following action in the extended theory
\begin{equation}
  \delta^\textrm{conformal}_\rho \varphi(x) = \left( \Delta - d \right) \langle \rho \rangle \,  \varphi(x).
\end{equation}
The volume-dependence can then be extracted using the canonical transformation $(\varphi,\pi_\varphi) \to (\bar\varphi, \bar\pi_\varphi)$:
\begin{equation}
    \varphi = \left( \frac{V}{V_0} \right)^{2 \left(\frac \Delta d - 1\right)}\bar\varphi,\,\,\,\,\,\,\,\,\,  \pi_\varphi = \left( \frac{V}{V_0} \right)^{-2 \left(\frac \Delta d - 1\right)}\bar\pi_\varphi.
\end{equation}

The volume expansion of the Shape Dynamics Hamiltonian can now be performed. In general, consistency of the equations will limit the possibilities for $U(\varphi)$ for a particular anomalous scaling, $\Delta$. Because the volume expansion will, in general, depend on $\varphi$ and $\pi_\varphi$, the volume expansion of the Hamilton--Jacobi equation will be modified. This will, in turn, affect the expansion coefficients of the volume expansion.

For a simple illustration of this, consider the case where $\Delta = d$. In this case, $\varphi$ has no conformal scaling. The zeroth-order equation is
\begin{equation}
    H_\textrm{(0)} = -\left( \frac{d(d-1)sk}{l^2} + \frac{d}{4(d-1)}P^2 + U(\varphi_{(0)}) \right),
\end{equation}
which is only consistent and non-trivial (i.e., $\varphi \neq \textrm{const}$) if is the scalar field is free so that $U(\varphi) = 0$. We simplify the calculation with the gauge choice
\begin{equation}
    \bar R - \bar h_{ab} \bar\nabla_a \varphi \bar\nabla_b \varphi = \textrm{const} \equiv \tilde R_\varphi.
\end{equation}
Then, the first order equations lead to
\begin{equation}
    H_{(1)} = s \tilde R_\varphi,\,\,\,\,\,\,\,\,\,\, \omega_{(1)} = 1.
\end{equation}
Subsequent orders will, therefore, be unchanged from the results obtained in the pure gravity case with $\tilde R \to \tilde R_\varphi$ until $n = d$. At order $n=d$, we obtain:
\begin{equation}
    H_{(d)} = -\frac{2s}{d} \left( \tilde R + (d-1) \tilde \nabla^2 \right)  \omega_{(d-1)} + \frac{{{\tilde \sigma}^a}_b \, {{\tilde \sigma}^b}_a + \pi_\varphi^2}{|\tilde h|} \,,
\end{equation}
taking the mean, we get
\begin{equation}
    H_{(d)} = \left\langle\frac{{{\tilde \sigma}^a}_b \, {{\tilde \sigma}^b}_a + \pi_\varphi^2}{|\tilde h|}\right\rangle \,.
\end{equation}
Using the substitution
\begin{equation}
    \pi_\varphi(x) \to \frac{\delta S}{\delta \varphi(x)},
\end{equation}
the volume expansion of the Hamilton--Jacobi equation becomes
\begin{equation}
   \begin{array}{rcl}
    0 &=&-\left( \frac{d(d-1)sk}{\ell^2} + \frac d {4(d-1)} \left(\frac{\delta S}{\delta V} \right)^2  \right) + s \tilde R_\varphi \left(\frac{V_0}{V}\right)^{2/d} +\\  && ~~ \left\langle\frac{\delta S}{\delta \tilde h_{ab}}\frac{\delta S}{\delta \tilde  h^{ab}} + \left(\frac {\delta S} {\delta \varphi} \right)^2\right\rangle\left(\frac{V_0}{V}\right)^2 + ...\, .
   \end{array}
\end{equation}
It is clear that we can still use the ansatz $S_{(0)} = \textrm{const}$ of homogeneous asymptotics to seed a recursion relation for the general volume expansion of $S$. However, because of the $\varphi$-dependence of $\tilde R_\varphi$ and the $\frac{\delta S}{\delta \varphi}$ term, the higher order expansion coefficients  will depend upon $\varphi$ as we intended to show. The explicit solution of $S$ for different matter fields is currently being investigated.

\end{appendix}


\clearpage

\providecommand{\href}[2]{#2}\begingroup\raggedright\endgroup

\end{document}